\documentclass[10pt,twocolumn,letterpaper]{article}

\usepackage{iccv}
\usepackage{times}
\usepackage{epsfig}
\usepackage{graphicx}
\usepackage{amsmath, fixmath}
\usepackage{amssymb}
\usepackage{multirow}
\usepackage{booktabs}
\usepackage{array}
\usepackage[dvipsnames]{xcolor}
\usepackage{authblk}
\usepackage{babel}
\usepackage{colortbl}

\iccvfinalcopy 

\begin{document}

\title{3D Medical Point Transformer: Introducing Convolution to Attention Networks for Medical Point Cloud Analysis}

\author[1]{Jianhui Yu}
\author[1]{Chaoyi Zhang}
\author[1]{Heng Wang}
\author[1]{Dingxin Zhang}
\author[2]{Yang Song}
\author[1]{Tiange Xiang}
\author[1]{\\Dongnan Liu}
\author[1]{Weidong Cai}
\affil[1]{School of Computer Science, University of Sydney, Australia}
\affil[2]{School of Computer Science and Engineering, University of New South Wales, Australia}


\maketitle
\ificcvfinal\thispagestyle{empty}\fi

\begin{abstract}
General point clouds have been increasingly investigated for different tasks, and recently Transformer-based networks are proposed for point cloud analysis.
However, there are barely related works for medical point clouds, which are important for disease detection and treatment.
In this work, we propose an attention-based model specifically for medical point clouds, namely 3D medical point Transformer (3DMedPT), to examine the complex biological structures.
By augmenting contextual information and summarizing local responses at query, our attention module can capture both local context and global content feature interactions.
However, the insufficient training samples of medical data may lead to poor feature learning, so we apply position embeddings to learn accurate local geometry and Multi-Graph Reasoning (MGR) to examine global knowledge propagation over channel graphs to enrich feature representations.
Experiments conducted on IntrA dataset proves the superiority of 3DMedPT, where we achieve the best classification and segmentation results.
Furthermore, the promising generalization ability of our method is validated on general 3D point cloud benchmarks: ModelNet40 and ShapeNetPart.
Code\footnote[1]{https://3dmedpt.github.io/} is released.
\end{abstract}

\section{Introduction}
Analysis of 3D point clouds has recently attracted much attention. Unlike 2D data structures that have a regular layout of pixels, 3D points are unordered and irregular, making it difficult to infer the underlying information. Efforts have been devoted to solving the problem by converting the 3D points to structured data, so that conventional methods can be applied.
Methods such as \cite{wu20153d, qi2016volumetric, maturana2015voxnet, xie2016deepshape, wu2016learning, su2015multi} regularize the structure of 3D points by voxelization or projection, which cause the loss of intrinsic geometric information.
\begin{figure}
\centering\includegraphics[width=\linewidth]{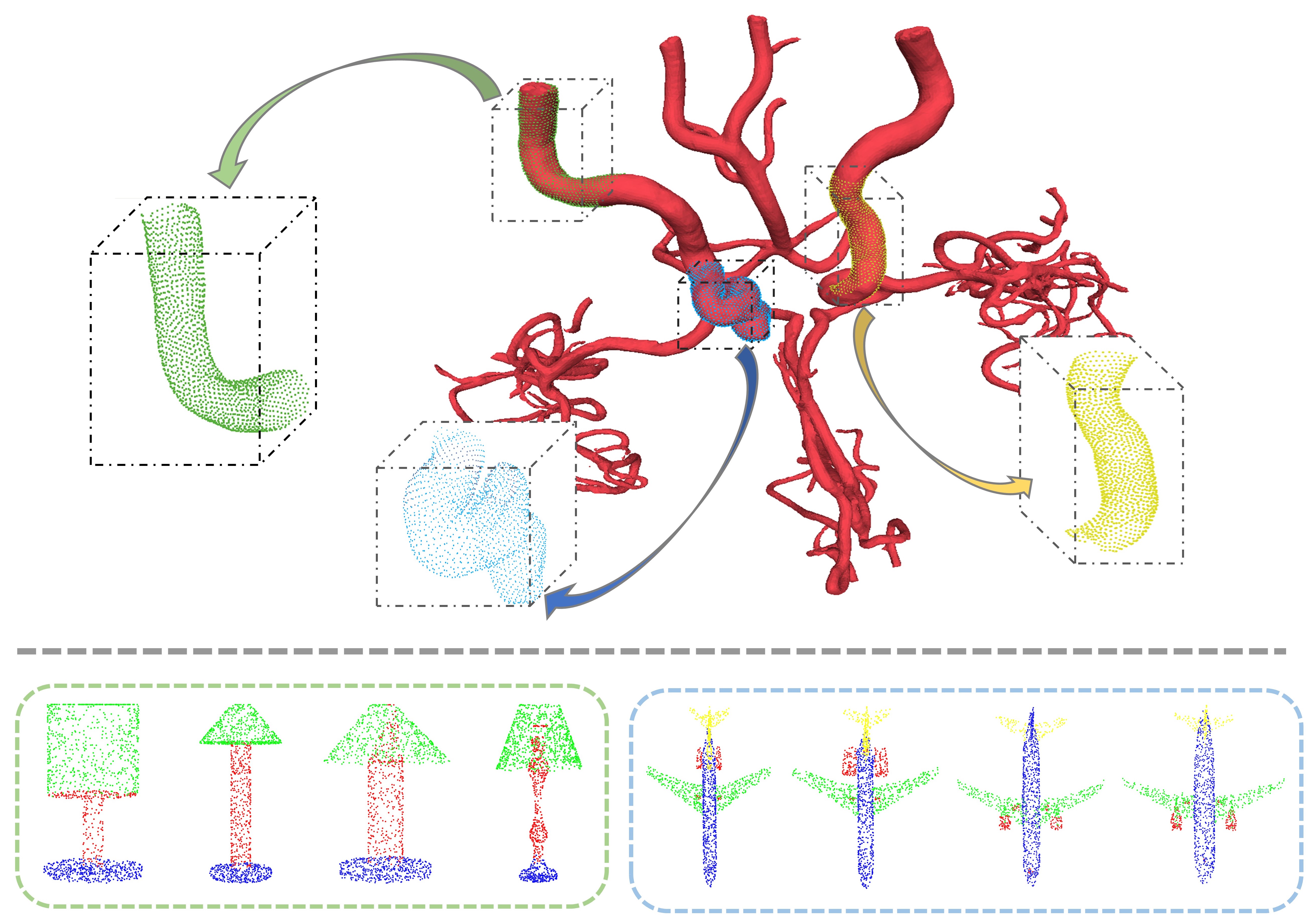}
\caption{
Top part: 3D medical point clouds in IntrA \cite{yang2020intra} with complex and diverse topology.
Bottom part: 3D general point clouds in ModelNet40 \cite{xie2016deepshape} with informative semantic structures and symmetry.} \label{fig:data_diff}
\end{figure}
Direct point-based approaches have thus been investigated. Encouraged by PointNet \cite{qi2017pointnet}, studies such as \cite{qi2017pointnet2,rsconv,pointgnn,pointcnn,Xiang_2021_ICCV,Zhang_2021_CVPR} directly learn spatial point features from 3D Euclidean space.

Although existing models can perform well on general 3D datasets, they could sometimes fail on medical data due to the domain gap.
Accurate pathological segments of medical data are important for disease diagnosis and treatment.
Nonetheless, 3D medical data may contain incomplete pathological structures, which are hard to be distinguished from healthy parts within one object \cite{yang2020intra}.
In addition, as shown in bottom part of Fig. \ref{fig:data_diff}, different types of the same object class usually have the same pattern for non-medical datasets.
In contrast, medical point clouds of the same object class have more diverse and complex shapes in geometry and topology \cite{medmeshcnn}, which introduce difficulties for object classification or segmentation.
The insufficient samples of medical data also make it difficult to learn distinctive shape descriptors.
Hence, it is essential to design an effective deep learning method which can demonstrate good performances on medical and also generalize well on non-medical data.

The recent success of Transformer in natural language processing (NLP) \cite{bert,xlnmet,vaswani2017attention} and vision analysis \cite{img16,kolesnikov2019big,detr,Vit} inspires us to investigate its potentials for the medical point cloud processing.
Self-attention is inherently order-invariant because attentional weights between the query and key remain the same if the input order is changed, which introduces permutation invariance, making it suitable for handling 3D point clouds.
Moreover, attention can model long-range dependencies and learn expressive features.
Based on these perspectives, we decide to analyze the medical point clouds based on attention layers.
To address the issue of the irregular layout intrinsic to medical data, we augment the input feature with contexts from local neighborhood before each attention module, making our network context-aware.
Meanwhile, the contextual information at query is further summarized via convolution to generate holistic geometric features.
Moreover, due to the small-scale medical dataset compared to general ones, we learn diverse positional embeddings at query, key and value, and we propose Multi-Graph Reasoning to concurrently establish multiple channel graphs over the same feature nodes, with variant learnable adjacency matrices to enrich feature expressions.


Our contributions are summarized as follows:
\textbf{(1)} We propose a Transformer-based network, namely 3DMedPT, to capture local context interactions via attention, and introduce convolution into Transformer to summarize global point features to obtain global content exchange within medical point clouds.
\textbf{(2)} We apply positional embeddings to address the irregular geometries of medical data.
\textbf{(3)} We design Multi-Graph Reasoning which captures global relations among feature channels to enrich the representational power of medical features at deep layers.
\textbf{(4)} Our model ranks the 1st in both classification and segmentation tasks in IntrA benchmark and reveals good generalization ability on ModelNet40 and ShapeNetPart.

\section{Related Work}
\noindent \textbf{3D Point Cloud Processing.}
Medical data play an important role in the development of human health research.
However, the number of datasets for 3D medical point clouds is quite small due to the expensive cost of data collection, thus existing methods are mostly proposed for general 3D objects.
Methods such as \cite{qi2017pointnet, qi2017pointnet2, densepoint} apply point-wise MLPs to analyze the irregular points.
PointNet \cite{qi2017pointnet} as a pioneer processes point clouds with shared MLPs and further aggregates the information via pooling layers, but local relations of geometric shapes are ignored.
Later works \cite{qi2017pointnet2, GDANet, xu2020geometry, Xiang_2021_ICCV} investigate the underlying geometry by grouping information from local ranges.
Some works \cite{pointcnn, wu2019pointconv, thomas2019kpconv, wang2018deep} handle point clouds with continuous convolutional kernels, while others \cite{wang2019dynamic, grid-gcn, graphattn} utilize graphs and graph convolution for point-wise feature encoding to learn local 3D structures.
However, these methods all adopt MLPs to process point features, which constrains the model ability in capturing more expressive shape information.

For medical point cloud processing, works such as \cite{bizjak2020vascular, he2020learning} directly borrow PointNet/PointNet++ for vessel labeling and aneurysm segmentation, and \cite{astolfi2020tractogram} modifies PointNet with a hand-engineered geometry learning algorithm.
Nonetheless, these works are still MLP-based, and the issue of complex and incomplete geometry of medical data cannot be well addressed.
In contrast, we process medical point clouds based on Transformer network in an effective manner.

\begin{figure*}
\centering\includegraphics[width=\textwidth]{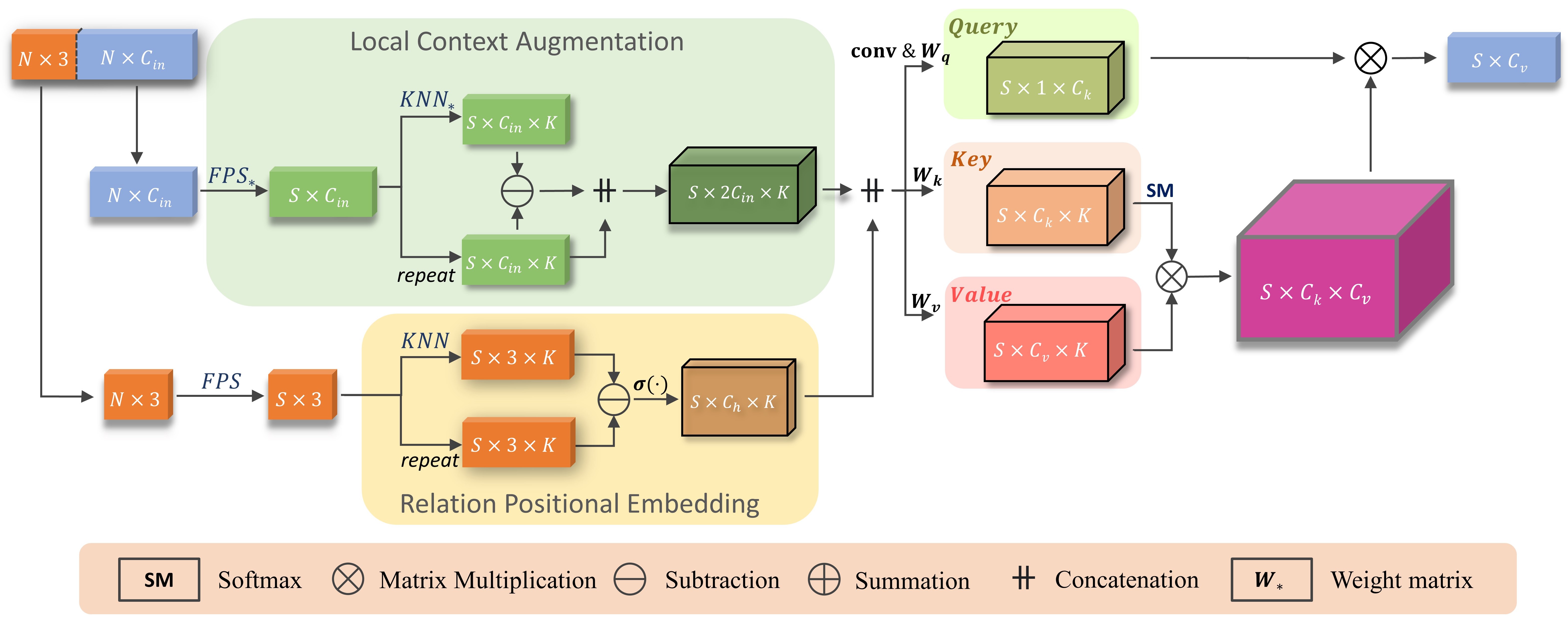}
\caption{
    Detailed architecture of our attention module in 3DMedPT, where KNN and FPS denote k-nearest neighbor and farthest point sampling operations \cite{qi2017pointnet2}, respectively.
} \label{fig:detailed_model}
\end{figure*}

\noindent \textbf{Transformer and MLP-like Methods.}
Transformer \cite{vaswani2017attention} has shown great success in NLP and machine translation tasks \cite{bert,xlnmet,vaswani2017attention}, which also encourages popular applications for 2D image processing \cite{img16,kolesnikov2019big,detr,Vit, liu2021Swin}.
However, due to the quadratic computational cost of attention maps, huge memory space will be taken to deal with even short input sequences.
A number of methods have been devoted to designing efficient attention implementations. \cite{roy2020efficient, child2019generating, kitaev2020reformer} use sparse matrix with strict constraints for efficient attention computation. Others \cite{performer, lambdanetworks, katharopoulos2020, axial} employ kernel factorization or matrix factorization to reduce the computational overhead.
Lambda attention \cite{lambdanetworks} reinterprets the attention as similarity kernels so that linear computations of attention are achieved, and axial-attention \cite{axial} decomposes 2D attention matrix into two 1D matrices along the width and height dimensions.
Recently, the Swin Transformer \cite{liu2021Swin} uses shifted windows to save computational cost.
In this work, we adopt the idea from Lambda attention by using linear functions for efficient computations.
However, different from the initial purpose of Lambda attention in the 2D domain, we modify input features by augmenting local contexts and relative positional bias to address the complex topology and irregular geometries of 3D medical point clouds.

Most recently, several Transformer-based networks have been proposed for general 3D point cloud analysis.
PCT \cite{guo2020pct} utilizes input embedding and offset attention to improve the network behavior.
Point Transformer \cite{pt} modifies vector attention with relative positional embeddings to construct hierarchical attention layers for point cloud analysis.
In this work, we propose an attention-based model that specifically works for medical point clouds with good generalization ability on non-medical datasets as well. 

There also exist works purely based on MLPs to truly explore the strength of MLPs.
MLP-mixer \cite{mlp-mixer} and Synthesizer \cite{tay2021synthesizer} are two pioneers in this stream that are solely built based on MLPs.
Variants such as \cite{chen2021cyclemlp,yu2021s,touvron2021resmlp} all focus on 2D domain.
PointMixer \cite{choe2021pointmixer} is the first work that uses MLP-like structures for point cloud analysis, where the query point feature is aggregated and updated from different dimensions.
However they do not perform quite well on classification task.
In contrast, our work reveals good behaviors on both medical and general datasets.

\noindent \textbf{Introducing Convolution to Transformers.}
Convolutions have been used to change the Transformer block in NLP and 2D image recognition, either by replacing multi-head attentions with convolution \cite{wu2019pay} or adding more convolution layers to capture local correlations \cite{wu2020lite, liu2020convtransformer, wu2021cvt}.
Different from all the previous works, we propose convolution operation (i.e., EdgeConv \cite{wang2019dynamic}) solely on query features to summarize local responses from unordered 3D points to generate global geometric representations, of which the purpose is totally opposite to \cite{liu2020convtransformer, wu2021cvt}.

\noindent \textbf{Graph-based Reasoning.}
Recently, graph convolutions \cite{gcn} have been adopted to capture relations between objects.
Graph-based reasoning has been adopted to achieve global relation reasoning over 2D image graphs \cite{latentgnn}.
However, general graph convolutions can only be safely applied when node connections are known, reasoning over point clouds is challenging since no link information between nodes is present.
\cite{superpoint} constructs graphs over 3D points with huge computational costs.
Inspired by \cite{ma2020global}, we construct graphs on feature channels to avoid dealing with a large number of points.
Moreover, to address the insufficient training samples in medical domain, we propose to construct multiple reasoning graphs over the same point features in parallel with learnable adjacency matrices for enriched global information learning.

\section{Method}
In this section, we firstly introduce the basic attention and Lambda attention (Sec.~\ref{sec:3.1}), then we explain how to embed local contexts for unordered 3D point clouds via contextual information augmentation (Sec.~\ref{neighbor}) and how to summarize local responses at query (Sec.~\ref{conv}).
Relative positional embeddings are proposed to handle local geometry of medical data (Sec.~\ref{relative_pos}).
Lastly, we propose MGR on feature channel domains to enrich the representations of learned features (Sec.~\ref{MGR}).

\subsection{Preliminary} \label{sec:3.1}
\noindent \textbf{Original Attention.}
Suppose we are given an unordered set of $N$ point features $\mathbf{P} \in \mathbb{R}^{N \times C}$ with $C$ feature channels. Three matrices can be learned through linear mappings:
\begin{equation} \label{qkv}
\mathbf{Q},\mathbf{K},\mathbf{V} = \mathbf{P}\mathbf{W}_{q},\mathbf{P}\mathbf{W}_{k},\mathbf{P}\mathbf{W}_{v},
\end{equation}
where $\mathbf{Q},\mathbf{K} \in \mathbb{R}^{N \times C_k}$ and $\mathbf{V} \in \mathbb{R}^{N \times C_v}$ have output feature dimensions of $C_k$ and $C_v$, respectively, and $W_{q}$, $W_k \in \mathbb{R}^{C \times C_k}$, and $W_v \in \mathbb{R}^{C\times C_v}$ are learnable weights.
Following the terms defined in \cite{vaswani2017attention}, we denote $\mathbf{Q},\mathbf{K},\mathbf{V}$ as query, key, and value, respectively, and the self-attention operation is formulated as follows:
\begin{equation} \label{norm_attn}
\text{Attn}(\mathbf{Q},\mathbf{K},\mathbf{V}) = \text{softmax}\left(\frac{\mathbf{Q}\mathbf{K}^{\top}}{\sqrt{C_k}}\right)\mathbf{V}.
\end{equation}
As shown in Eq.~\ref{norm_attn}, global attention weights calculated from $\mathbf{Q}$ and $\mathbf{K}$ have a time complexity $\mathcal{O}(N^{2}C_v)$ and space complexity $\mathcal{O}(N^2 + NC_k+NC_v)$, which increase quadratically when $N$ increases and consumes much computational resources.
Recently, existing works \cite{performer, lambdanetworks, katharopoulos2020} use linear operations to factorize the naive attention for computation acceleration.
In our design, we directly borrow the attention operation from Lambda attention \cite{lambdanetworks} as improving the attention efficiency is not our main focus.

\noindent \textbf{Lambda Attention.}
With the assistance of linear attention \cite{katharopoulos2020,performer} and kernel factorization, an efficient algorithm named as Lambda attention is proposed to bypass the need of materializing the attention maps \cite{lambdanetworks}. This process can be briefly expressed as:
\begin{equation} \label{new_attn}
\text{Attn}(\mathbf{Q},\mathbf{K},\mathbf{V}) = \mathbf{Q} {\left(\text{softmax}({\mathbf{K}})^{\top}\mathbf{V}\right)},
\end{equation}
where \textit{keys} are normalized through the softmax function, and $\text{softmax}({\mathbf{K}})^{\top}\mathbf{V} \in \mathbb{R}^{C_k \times C_v}$ is termed as content lambda \cite{lambdanetworks}, where each query can interact with the content lambda in a linear form.
Therefore, the time and space complexities are $\mathcal{O}(NC_{k}C_v)$ and $\mathcal{O}(NC_k+NC_v+C_{k}C_{v})$, respectively, where the computational cost could be largely reduced when $C_k \ll N$.
For efficient computation, we choose Lambda attention as our baseline model for medical point cloud processing.

\subsection{Local Context Augmentation} \label{neighbor}
Although self-attention is able to model long-range dependencies over the global domain, it cannot aggregate local information, which is essential in point cloud analysis \cite{guo2020pct}.
However, different from regular layouts such as 2D images where spatially neighboring pixels usually have high semantic correlations, 3D point clouds are unordered and nearby points can have no geometric or semantic relations due to permutation variance.
Hence, instead of using local attention \cite{StandAlone} which may constrain the model's receptive field, we reform the input feature before each attention layer by defining a local context region, with the assumption that spatially closed points in Euclidean coordinates can have some relations for geometric study.
We thus follow the idea of PointNet++ \cite{qi2017pointnet2} by firstly downsampling the points using farthest point sampling (FPS) and then group features from local contexts as DGCNN \cite{wang2019dynamic}.

Specifically, given $xyz$ coordinates of $N$ input points $\mathbf{P}=\{\mathbf{p}_i \in \mathbb{R}^{3}\}^{N}_{i=1}$ and the corresponding features $\mathbf{F}=\{\mathbf{f}_i \in \mathbb{R}^{C_{in}}\}^{N}_{i=1}$ with $C_{in}$ feature channels, the whole process can be formulated as follows:
\begin{equation} \label{eq:knn}
    \begin{gathered}
        \mathcal{N}(\mathbf{p}_i) = \text{KNN}(\mathbf{P}, ||\mathbf{p}_i - \mathbf{p}_j||_{2}^{2}), \; \mathbf{p}_j \in \mathbf{P}_{S}, \\
        \mathbf{f}_{i}^{\prime} = [\mathbf{f}_j - \mathbf{f}_i, \mathbf{f}_i]_{j \in \mathcal{N}(\mathbf{p}_i)} \in \mathbb{R}^{K \times 2C_{in}},
    \end{gathered}
\end{equation}
where $\mathbf{P}_{S}$ is a point set downsampled from $\mathbf{P}$ using FPS, $\text{KNN}(\cdot)$ is K-nearest neighbor function, $[\cdot, \cdot]$ is concatenation, and $\mathbf{f}_{i}^{\prime}$ is the augmented feature with local contexts.
In this case, $\mathbf{f}_{i}^{\prime}$ can now retain the locality property.


\subsection{Convolution at Query} \label{conv}
This convolution operation proposed at query has two purposes: (1) it allows us to aggregate local responses and update geometric features at query; (2) Since global interaction should be considered between the query and content lambda $\text{softmax}({\mathbf{K}})^{\top}\mathbf{V}$, so it is natural to capture global features from query.
Therefore, we introduce convolution (i.e., EdgeConv) to Transformer to generate global query information.

Formally, the query tensor is obtained from the updated features $\mathbf{F}^{\prime}=\{\mathbf{f}^{\prime}_i\}^{N}_{i=1} \in \mathbb{R}^{N \times K \times 2C_{in}}$ by using EdgeCov:
\begin{equation}
    \mathbf{Q} = \operatorname{EdgeConv}(\mathbf{F}^{\prime})W_{q} \in \mathbb{R}^{N \times C_{k}},
\end{equation}
where $W_{q} \in \mathbb{R}^{2C_{in} \times C_{k}}$.
In this case, the local information is integrated into $\mathbf{Q}$, then we apply linear projections on flattened $\mathbf{F}^{\prime}$ to compute $\mathbf{K}$ and $\mathbf{V}$ as follows:
\begin{equation}
    \begin{gathered}
        \mathbf{K} = \operatorname{Flatten}(\mathbf{F}^{\prime})W_{k} \in \mathbb{R}^{(N \times K) \times C_{k}}, \\
        \mathbf{V} = \operatorname{Flatten}(\mathbf{F}^{\prime})W_{v} \in \mathbb{R}^{(N \times K) \times C_{v}},
    \end{gathered}
\end{equation}
where $W_{k} \in \mathbb{R}^{2C_{in} \times C_{k}}$ and $W_{v}\in \mathbb{R}^{2C_{in} \times C_{v}}$.
Based on the above modifications, we show an improved version of Eq. \ref{new_attn} in a per-point form as:
\begin{equation} \label{eq:ynew1}
    \mathbf{y}_{i} = \mathbf{q}_{i}\big(\text{softmax}(\mathbf{k}_{i})^{\top}\mathbf{v}_{i}\big),
\end{equation}
where $\mathbf{y}_{i} \in \mathbb{R}^{C_{v}}$ is the layer output, $\mathbf{k}_{i} \in \mathbb{R}^{K \times C_{k}}$ and $\mathbf{v}_{i} \in \mathbb{R}^{K \times C_{v}}$.



\begin{figure}
\centering\includegraphics[width=0.9\linewidth]{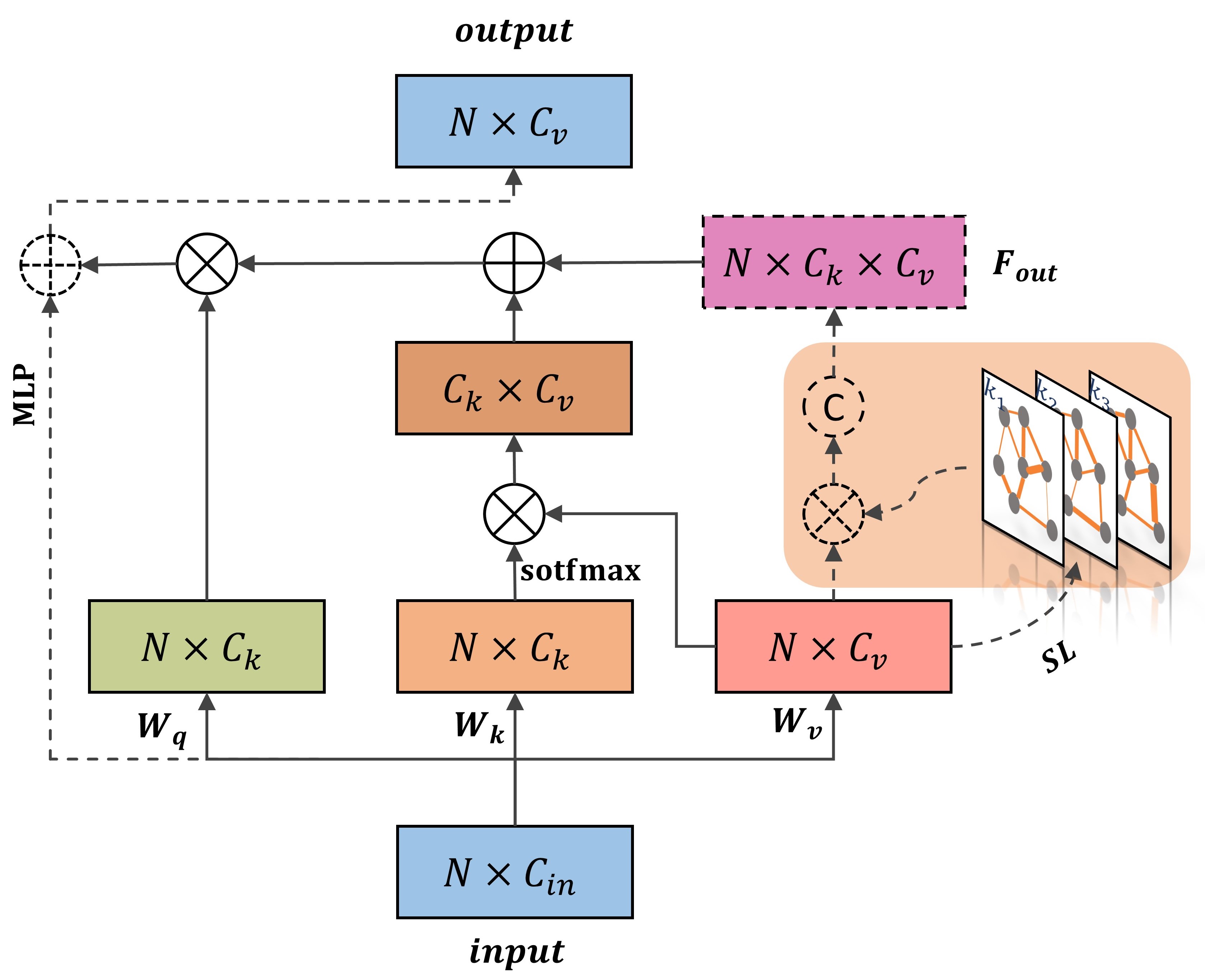}
\caption{Modified attention module with MGR, where SL denotes the self-loop and residual connection is applied to compensate the low-level information loss. All dotted components imply different designs with the original Lambda attention \cite{lambdanetworks}. 
} \label{fig:MGR}
\end{figure}

\begin{figure*}
\centering\includegraphics[width=\textwidth]{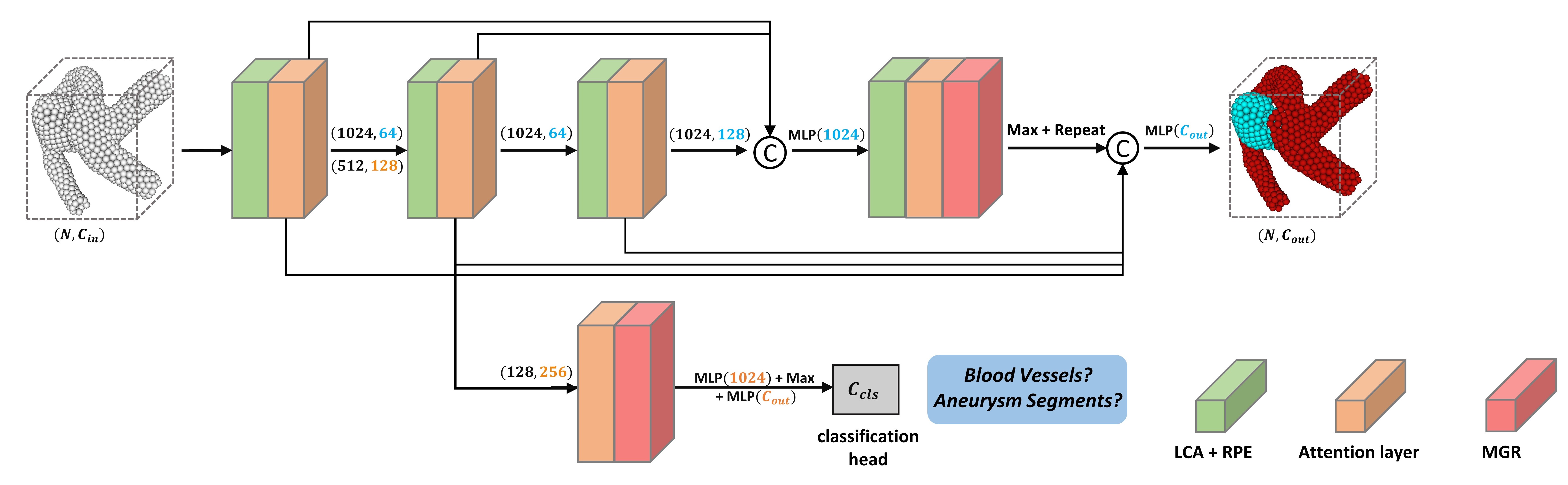}
\caption{
    The overall architecture of 3DMedPT for medical point cloud analysis.
    Numbers in black, \textcolor{Cerulean}{blue}, and \textcolor{BurntOrange}{orange} indicate the point number, the feature dimension for classification, and the feature dimension for segmentation.
    LCA and RPE denote local context augmentation and relative positional embedding, respectively.
} \label{fig:overall_model}
\end{figure*}

\subsection{Relative Positional Embedding} \label{relative_pos}
As mentioned in \cite{pt}, positional information is critical for 3D point cloud processing.
Especially for medical data where the structure is incomplete and complex, embedding positional bias encourages the model to focus on local geometry.
In this work, we use relative positional information since computing absolute positions requires storing an ordered list before and after point permutations, which increases the computational overhead.
In addition, we integrate the relative positions from local contexts with input feature as we empirically find that using addition cannot give the best result.
Therefore, we define learnable relative positional embedding as follows:
\begin{equation}
\begin{gathered}
    \mathbf{h}_{i} = \sigma\big([\mathbf{p}_j - \mathbf{p}_i]_{j \in \mathcal{N}(\mathbf{p}_{i})}\big) \in \mathbb{R}^{K \times C_{h}}, \\
    \mathbf{f}_{i}^{\prime} = [\mathbf{f}_{j}-\mathbf{f}_{i}, \mathbf{f}_{i}, \mathbf{h}_{i}]_{j \in \mathcal{N}(\mathbf{p}_{i})},
\end{gathered}
\end{equation}
where $\sigma(\cdot)$ is an MLP.

However, learning accurate positional bias is quite hard when dataset is too small \cite{med-transformer}, which is our case when dealing with medical data.
we therefore apply different MLPs $\sigma(\cdot)$ to learn positional information at query\&key\&value positions for accurate and complex medical geometry learning.
As illustrated in Fig. \ref{fig:detailed_model}, we improve Lambda attention by introducing positional bias terms at a modest cost to address the irregular geometric traits lurking inside medical data.


\subsection{Multi-Graph Reasoning} \label{MGR}
Feature representations are critical for model performance, especially when dealing with insufficient training samples, which is often the case in medical domain.
In our method, we propose to enrich the feature representations by exploiting global relations of content in deep layers of our attention block with a Multi-Graph Reasoning (MGR) module based on graph reasoning \cite{glore} and graph convolutions \cite{gcn}.
As suggested in \cite{ma2020global}, graphs can be constructed on feature channels by learning graph nodes from channels, which could save the computational cost for the case of large input numbers.
In contrast to \cite{ma2020global} that only a single graph is established over channels, we design an MGR module to initialize multiple graphs simultaneously with learnable adjacency matrices, enhancing the diversity of node features via various graph states.
As shown inside the pink box of Fig. \ref{fig:MGR}, MGR is adapted on \textit{values} to replace the original positional encoding part in the last attention layer, where neighboring information augmentation is intentionally ignored for global information aggregation and relational interactions.
Hence, the output $\mathbf{F}_{out}$ of MGR module can be formulated as:
\begin{equation} \label{eq:glore}
    \mathbf{F}_{out} = [\text{ReLU}\big((\mathbf{V}+\mathbf{I})\mathbf{a}_{i}\big)]_{i \in C_{k}}, \quad \mathbf{a}_{i} \in \mathbb{R}^{C_v \times C_v},
\end{equation}
where $\mathbf{I}$ is the identity matrix, indicating that the self-loop of each node is introduced. $C_k$ graphs and the corresponding adjacency matrices $\mathbf{a}_{i}$ are concurrently computed from $C_{v}$ channel nodes, such that contextual interactions between nodes can be modeled and structural relations are captured for feature learning.

\subsection{Implementation Details}
The overall model architecture for 3D object classification and part segmentation is shown in Fig. \ref{fig:overall_model}.
In our overall network model, the input feature dimension $C_{in}$ is set to 3 representing 3D normal vectors, and if there is no normal vectors, $C_{in}$ will be the $xyz$ positions.
The output feature dimension $C_{out}$ is set to 2 for classifying and segmenting blood vessels or aneurysms, with binary cross entropy as the loss function.
As shown in Fig. \ref{fig:overall_model}, we follow the architecture of PointNet++ for classification, where points are downsampled and features are embedded with locality, followed by the attention module to enhance interactions between local contexts and global contents.
MGR is applied to further enrich the feature representations with graph reasoning, and maxpooling is employed to summarize the information while being insensitive to permutation as a symmetric function.
For segmentation, we adopt a DGCNN-like network.
Each MLP block contains a linear mapping layer, batch normalization layer and ReLU.
To address bottleneck issues caused by a small $C_v$, the multi-query algorithm \cite{lambdanetworks} is utilized so that $h$ different query heads are constructed and concatenated, deriving a new output feature $\mathbf{y}_{i} \in \mathbb{R}^{hC_{v}}$.



\section{Experiments}
In this section, we present our experimental results on medical 3D points (IntrA \cite{yang2020intra}) and show the model performance on other medical datasets in the supplementary material.
We also test the generalization ability on non-medical 3D point clouds (ModelNet40 \cite{wu20153d} and ShapeNetPart \cite{shapenet}).
Extensive experiments for ablation studies are conducted as well.
The training details are reported in the supplementary material.

\subsection{3D Object Classification} \label{cls}
We first examine the model behavior on IntrA and then investigate our model on ModelNet40.

\noindent \textbf{IntrA.}
IntrA \cite{yang2020intra} is a 3D medical point cloud dataset for binary classification and part segmentation to distinguish blood vessels and aneurysms, which contains mesh and point representations of the data structure.
We use the point cloud with overall 2025 samples for classification.
Five-fold cross-validation was adopted with F1-score and per-class testing accuracy as evaluation metrics.

As shown in Table \ref{intra_cls}, our method reaches the highest accuracies of 94.06\% with 512 points for aneurysm (A.) and 99.24\% with 1024 points for blood vessel (V.) detection by using the PointNet++ backbone, which overpass the original work by 0.7\% and 6.3\% respectively. 
Our 3DMedPT achieves the best F1 score of 0.936 with 1024 points, which outperforms its Transformer counterpart PCT by 2.4\%.
It is also 3.3\% and 9.1\% higher than the latest attention-based works: PAConv \cite{paconv} and AdaptConv \cite{adapconv}, showing the superiority of our method on 3D medical point cloud analysis.

\begin{table} 
\small
\centering
\caption{Classification results of per-class accuracy and F1-score on healthy vessel segments (V.) and aneurysm segments (A.) with all input features. Results are averaged across all 5 folds.} \label{intra_cls}
\begin{tabular}{l|llll} 
\toprule
Method                    & \#Points & V. (\%) & A. (\%) & F1      \\ 

\hline
\multirow{3}{*}{PointNet \cite{qi2017pointnet}}   & \cellcolor{Apricot}512      & \cellcolor{Apricot}94.45   & \cellcolor{Apricot}67.66   & \cellcolor{Apricot}0.691  \\
                            & 1024     & 94.98   & 64.96   & 0.684  \\
                            & \cellcolor{Lavender}2048     & \cellcolor{Lavender}93.74   & \cellcolor{Lavender}69.50    & \cellcolor{Lavender}0.692  \\ 
\hline
\multirow{3}{*}{PointNet++ \cite{qi2017pointnet2}} & \cellcolor{Apricot}512      & \cellcolor{Apricot}98.52   & \cellcolor{Apricot}86.69   & \cellcolor{Apricot}0.893  \\
                            & 1024     & 98.52   & 88.51   & 0.903  \\
                            & \cellcolor{Lavender}2048     & \cellcolor{Lavender}98.76   & \cellcolor{Lavender}87.31   & \cellcolor{Lavender}0.902  \\
\hline
\multirow{3}{*}{PointCNN \cite{pointcnn}}   & \cellcolor{Apricot}512      & \cellcolor{Apricot}98.38   & \cellcolor{Apricot}78.25   & \cellcolor{Apricot}0.849  \\
                            & 1024     & 98.79   & 81.28   & 0.875  \\
                            & \cellcolor{Lavender}2048     & \cellcolor{Lavender}98.95   & \cellcolor{Lavender}85.81   & \cellcolor{Lavender}0.904  \\ 
\hline
\multirow{3}{*}{PointConv \cite{wu2019pointconv}}
                            & \cellcolor{Apricot}512      & \cellcolor{Apricot}99.21   & \cellcolor{Apricot}91.96   & \cellcolor{Apricot}0.915  \\
                            & 1024     & 98.89   & 83.57   & 0.883  \\
                            & \cellcolor{Lavender}2048     & \cellcolor{Lavender}98.61   & \cellcolor{Lavender}90.47   & \cellcolor{Lavender}0.883  \\ 
\hline
\multirow{3}{*}{SO-Net \cite{sonet}}     & \cellcolor{Apricot}512      & \cellcolor{Apricot}98.76   & \cellcolor{Apricot}84.24   & \cellcolor{Apricot}0.884   \\
                            & 1024     & 98.88   & 81.21   & 0.868  \\
                            & \cellcolor{Lavender}2048     & \cellcolor{Lavender}98.88   & \cellcolor{Lavender}83.94   & \cellcolor{Lavender}0.885   \\ 
\hline
\multirow{3}{*}{SpiderCNN \cite{xu2018spidercnn}}  & \cellcolor{Apricot}512      & \cellcolor{Apricot}98.05   & \cellcolor{Apricot}84.58   & \cellcolor{Apricot}0.869  \\
                            & 1024     & 97.28   & 87.90   & 0.872  \\
                            & \cellcolor{Lavender}2048     & \cellcolor{Lavender}97.28   & \cellcolor{Lavender}84.89   & \cellcolor{Lavender}0.866  \\ 
\hline
\multirow{3}{*}{DGCNN \cite{wang2019dynamic}}      & \cellcolor{Apricot}512      & \cellcolor{Apricot}95.22   & \cellcolor{Apricot}60.73   & \cellcolor{Apricot}0.658  \\
                            & 1024     & 95.34   & 72.21   & 0.738  \\
                            & \cellcolor{Lavender}2048     & \cellcolor{Lavender}97.93   & \cellcolor{Lavender}83.40    & \cellcolor{Lavender}0.859  \\ 
\hline
\multirow{3}{*}{GS-Net \cite{xu2020geometry}}
                            & \cellcolor{Apricot}512      & \cellcolor{Apricot}98.55   & \cellcolor{Apricot}83.84   & \cellcolor{Apricot}0.873  \\
                            & 1024     & 98.78   & 83.08   & 0.872  \\
                            & \cellcolor{Lavender}2048     & \cellcolor{Lavender}98.39   & \cellcolor{Lavender}85.74   & \cellcolor{Lavender}0.882  \\
\hline
\multirow{3}{*}{PCT \cite{guo2020pct}}
                            & \cellcolor{Apricot}512      & \cellcolor{Apricot}99.03   & \cellcolor{Apricot}89.07   & \cellcolor{Apricot}0.911  \\
                            & 1024     & 98.87   & 89.71   & 0.914  \\
                            & \cellcolor{Lavender}2048     & \cellcolor{Lavender}98.96   & \cellcolor{Lavender}89.49   & \cellcolor{Lavender}0.917  \\ 
\hline
\multirow{3}{*}{PAConv \cite{paconv}}
                            & \cellcolor{Apricot}512      & \cellcolor{Apricot}98.53   & \cellcolor{Apricot}89.00   & \cellcolor{Apricot}0.904  \\
                            & 1024     & 98.98   & 89.71   & 0.906  \\
                            & \cellcolor{Lavender}2048     & \cellcolor{Lavender}98.19   & \cellcolor{Lavender}85.74   & \cellcolor{Lavender}0.882  \\
\hline
\multirow{3}{*}{AdaptConv \cite{adapconv}}
                            & \cellcolor{Apricot}512      & \cellcolor{Apricot}97.58   & \cellcolor{Apricot}79.99   & \cellcolor{Apricot}0.809  \\
                            & 1024     & 99.05   & 82.90   & 0.858  \\
                            & \cellcolor{Lavender}2048     & \cellcolor{Lavender}97.87   & \cellcolor{Lavender}75.94   & \cellcolor{Lavender}0.799  \\ 
\hline
\multirow{3}{*}{3DMedPT}    & \cellcolor{Apricot}512        & \cellcolor{Apricot}99.02            & \cellcolor{Apricot}\textbf{94.06}     & \cellcolor{Apricot}0.920       \\
                            & 1024       & \textbf{99.24}   & 93.26              & \textbf{0.936}       \\
                            & \cellcolor{Lavender}2048       & \cellcolor{Lavender}99.07            & \cellcolor{Lavender}93.49              &  \cellcolor{Lavender}0.931      \\
\bottomrule
\end{tabular}
\end{table}

\noindent \textbf{ModelNet40.}
ModelNet40 dataset \cite{wu20153d} consists of 13,211 3D synthetic models for general objects, with 9843 training samples and 2468 testing samples ranged within 40 classes.
We uniformly sample 1024 points only with 3D coordinates as input features, and shuffle the points as in \cite{qi2017pointnet2}.
Our performance compared to other state-of-the-art methods is listed in Table \ref{table:modelnet40_table}.

We achieve an accuracy of 93.4\%, which overpasses most typical point-based designs \cite{xu2020geometry, wu2019pointconv, wang2019dynamic}.
Additionally, our model performance is better than some networks based on attention algorithms (e.g., Set Transformer \cite{lee2019set}, PAT \cite{pat} and Point2Sequence \cite{p2seq}), and our model is also better than the Transformer counterpart PCT \cite{guo2020pct} with a much smaller model size (see Table \ref{tab:model_eff}).
Our classification accuracy is lower than recently proposed Point Transformer \cite{pt} by only 0.3\%, while this small gap validates the good generalization ability of 3DMedPT. Hence, our design can not only deal with medical dataset with complex topology such as blood vessels or aneurysms, but also distribute the excellence to regular 3D shapes.

\begin{table} 
\small
\centering
\caption{Classification results on ModelNet40 with different input types and point numbers.
} \label{table:modelnet40_table}
\begin{tabular}{l|lll} 
\toprule
Method       & Input      & \#Points & Acc. (\%)  \\ 
\hline
Set Transformer \cite{lee2019set}    & xyz & 5k       & 90.4       \\
PointCNN \cite{pointcnn}     & xyz        & 1k       & 91.7       \\
DGCNN \cite{wang2019dynamic}         & xyz        & 1k       & 92.2       \\
Point2Sequence \cite{p2seq}   & xyz & 1k       & 92.6       \\
GS-Net \cite{xu2020geometry}        & xyz        & 1k       & 92.9      \\
RS-CNN \cite{rsconv}        & xyz        & 1k       & 92.9      \\
SO-Net \cite{sonet}       & xyz        & 2k       & 90.9       \\
KPConv \cite{thomas2019kpconv}        & xyz        & 7k       & 92.9       \\ 
PCT \cite{guo2020pct}    & xyz  & 1k       & 93.2       \\
AdaptConv \cite{adapconv}           & xyz & 1k & 93.4 \\
PAConv \cite{paconv}                & xyz & 1k & 93.6 \\
Point Transformer \cite{pt}    & xyz & 1k       & \textbf{93.7}        \\
\hline
PAT \cite{pat}          & xyz + norm & 1k       & 91.7       \\
PointConv \cite{wu2019pointconv}     & xyz + norm & 1k       & 92.5       \\
PointASNL \cite{pointasnl}    & xyz + norm & 1k       & 93.2       \\
PointNet++ \cite{qi2017pointnet2}   & xyz + norm & 5k       & 91.9       \\
SpiderCNN \cite{xu2018spidercnn}     & xyz + norm & 5k       & 92.4       \\ 
\hline
3DMedPT         & xyz        & 1k       & 93.4       \\
\bottomrule
\end{tabular}
\end{table}

\subsection{3D Part Segmentation} \label{seg}
We then validate the segmentation ability of our model on both IntrA and ShapeNetPart, with the same data augmentation method as Sec. \ref{cls}.

\noindent \textbf{IntrA.}
There are a total of 116 annotated samples for part segmentation task in IntrA, where the boundary lines are grouped into aneurysm segments, making it a binary segmentation task.
Five-fold cross-valuation is still applied with evaluation metrics based on Point Intersection over Union (IoU) and Sørensen–Dice cefficient (DSC).
Results are reported in Table \ref{tab:seg}.
It can be seen that we achieve the highest IoU and DSC values of 94.82\% and 97.29\% for parent vessels segmentation with 512 input points.
Meanwhile, our 3DMedPT also has the best performance on the aneurysm segmentation with 1024 points, resulting in IoU and DSC values of 82.39\% and 89.71\%.
Our work outperforms PAConv and AdaptConv by a large margin by 4.7\% and 9.5\% on A. IoU and 2.8\% and 4.2\% on V. IoU, which exhibits our superiority on medical data.

\begin{figure}
\centering
\includegraphics[width=\linewidth]{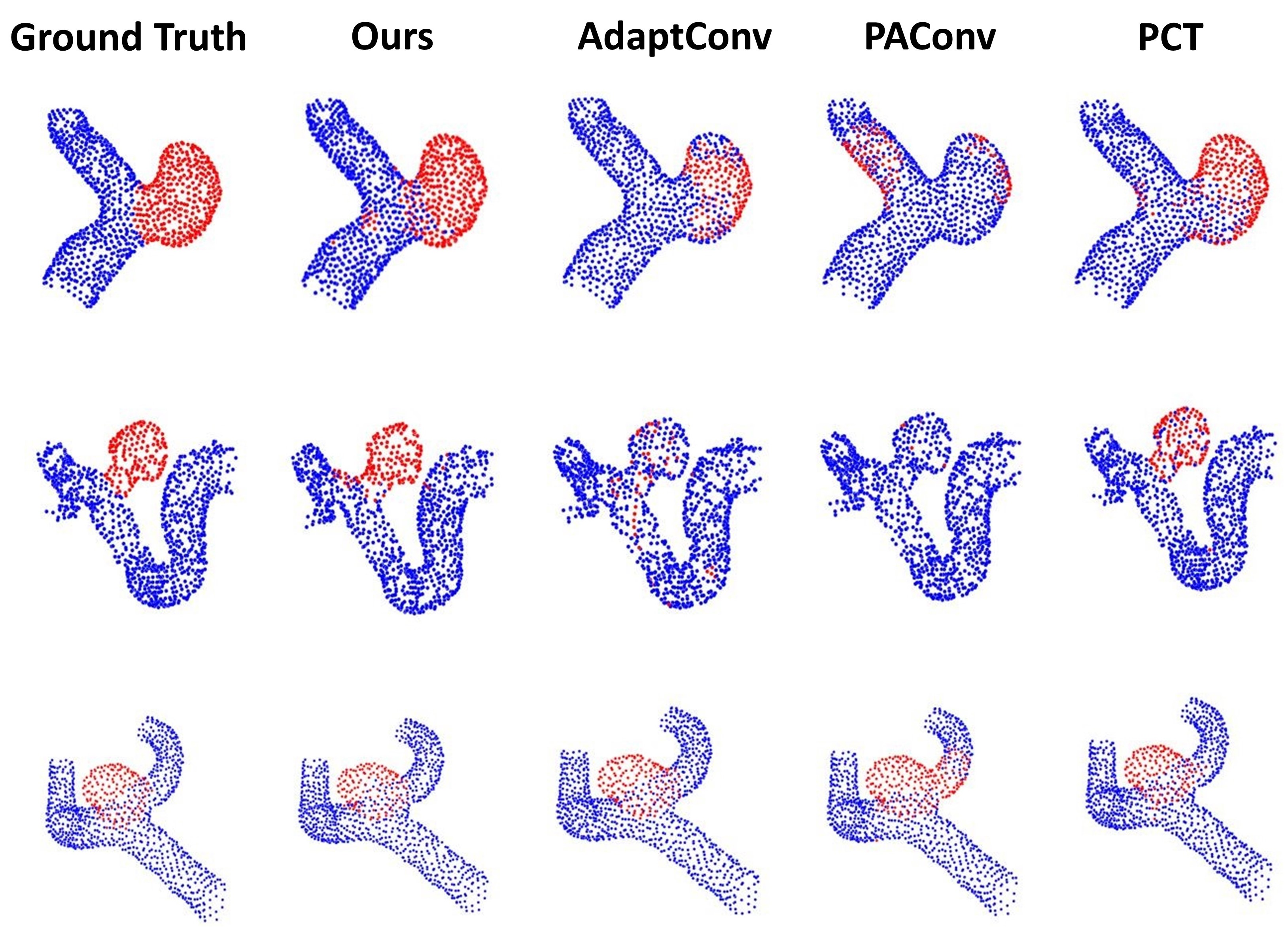}
\caption{Qualitative comparisons on IntrA segmentation. Ground-truth samples are shown in the first column for reference.} \label{fig:seg_result}
\end{figure}

To further examine the model behavior, we qualitatively evaluate our approach with respect to some recent works such as PAConv.
Ground truth samples are shown in the 1st column for reference in Fig. \ref{fig:seg_result}.
As can be seen, when the aneurysm takes a large size ratio of the blood vessel (row 1), our model performs the best and PCT cannot fully understand the shape of aneurysm, while PCT gives the similar segmentation results as ours in other cases (rows 2-3).
However, we can see that the latest work PAConv totally fails when complicated structures are encountered (row 2).

\begin{table}
\small
\centering
\caption{Segmentation results of each point-based network. V. and A. represent parent vessel segments and aneurysm segments.} \label{tab:seg}
\begin{tabular}{l|lllll} 
\toprule
\multirow{2}{*}{Method} & \multicolumn{1}{l}{\multirow{2}{*}{\#Points}} & \multicolumn{2}{c}{IoU (\%)} & \multicolumn{2}{c}{DSC (\%)} \\
& \multicolumn{1}{l}{} & V. & \multicolumn{1}{l}{A.} & V. & A. \\ 
\hline
\multirow{3}{*}{PointNet \cite{qi2017pointnet}}     & \cellcolor{Apricot}512  & \cellcolor{Apricot}73.99 & \cellcolor{Apricot}37.30 & \cellcolor{Apricot}84.05 & \cellcolor{Apricot}48.96 \\
                                                    & 1024 & 75.23 & 37.07 & 85.00 & 48.38 \\
                                                    & \cellcolor{Lavender}2048 & \cellcolor{Lavender}74.22 & \cellcolor{Lavender}37.75 & \cellcolor{Lavender}84.17 & \cellcolor{Lavender}49.59 \\ 
\hline
\multirow{3}{*}{PointNet++ \cite{qi2017pointnet2}}  & \cellcolor{Apricot}512  & \cellcolor{Apricot}93.42 & \cellcolor{Apricot}76.22 & \cellcolor{Apricot}96.48 & \cellcolor{Apricot}83.92 \\
                                                    & 1024 & 93.35 & 76.38 & 96.47 & 84.62 \\
                                                    & \cellcolor{Lavender}2048 & \cellcolor{Lavender}93.24 & \cellcolor{Lavender}76.21 & \cellcolor{Lavender}96.40 & \cellcolor{Lavender}84.64 \\ 
\hline
\multirow{3}{*}{PointCNN \cite{pointcnn}}           & \cellcolor{Apricot}512  & \cellcolor{Apricot}92.49 & \cellcolor{Apricot}70.65 & \cellcolor{Apricot}95.97 & \cellcolor{Apricot}78.55 \\
                                                    & 1024 & 93.47 & 74.11 & 96.53 & 81.74 \\
                                                    & \cellcolor{Lavender}2048 & \cellcolor{Lavender}93.59 & \cellcolor{Lavender}73.58 & \cellcolor{Lavender}96.62 & \cellcolor{Lavender}81.36 \\ 
\hline
\multirow{3}{*}{SO-Net \cite{sonet}}                & \cellcolor{Apricot}512  & \cellcolor{Apricot}94.22 & \cellcolor{Apricot}80.14 & \cellcolor{Apricot}96.95 & \cellcolor{Apricot}87.90 \\
                                                    & 1024 & 94.42 & 80.99 & 97.06 & 88.41 \\
                                                    & \cellcolor{Lavender}2048 & \cellcolor{Lavender}94.46 & \cellcolor{Lavender}81.40 & \cellcolor{Lavender}97.09 & \cellcolor{Lavender}88.76 \\ 
\hline
\multirow{3}{*}{SpiderCNN \cite{xu2018spidercnn}}   & \cellcolor{Apricot}512  & \cellcolor{Apricot}90.16 & \cellcolor{Apricot}67.25 & \cellcolor{Apricot}94.53 & \cellcolor{Apricot}75.82 \\
                                                    & 1024 & 87.95 & 61.60 & 93.24 & 71.08 \\
                                                    & \cellcolor{Lavender}2048 & \cellcolor{Lavender}87.02 & \cellcolor{Lavender}58.32 & \cellcolor{Lavender}92.17 & \cellcolor{Lavender}67.74 \\ 
\hline
\multirow{3}{*}{PointConv \cite{wu2019pointconv}}   & \cellcolor{Apricot}512  & \cellcolor{Apricot}94.16 & \cellcolor{Apricot}79.09 & \cellcolor{Apricot}96.89 & \cellcolor{Apricot}86.01 \\
                                                    & 1024 & 94.59 & 79.42 & 97.15 & 86.29 \\
                                                    & \cellcolor{Lavender}2048 & \cellcolor{Lavender}94.65 & \cellcolor{Lavender}79.53 & \cellcolor{Lavender}97.18 & \cellcolor{Lavender}86.52 \\ 
\hline
\multirow{3}{*}{GS-Net \cite{xu2020geometry}}       & \cellcolor{Apricot}512  & \cellcolor{Apricot}90.06 & \cellcolor{Apricot}64.48 & \cellcolor{Apricot}94.62 & \cellcolor{Apricot}74.54 \\
                                                    & 1024 & 90.93 & 66.29 & 95.10 & 78.85 \\
                                                    & \cellcolor{Lavender}2048 & \cellcolor{Lavender}91.06 & \cellcolor{Lavender}65.76 & \cellcolor{Lavender}95.15 & \cellcolor{Lavender}75.06 \\ 
\hline
\multirow{3}{*}{PCT \cite{guo2020pct}}              & \cellcolor{Apricot}512  & \cellcolor{Apricot}92.49 & \cellcolor{Apricot}78.09 & \cellcolor{Apricot}96.08 & \cellcolor{Apricot}85.84 \\
                                                    & 1024 & 92.05 & 78.12 & 95.85 & 86.77 \\
                                                    & \cellcolor{Lavender}2048 & \cellcolor{Lavender}91.66 & \cellcolor{Lavender}77.10 & \cellcolor{Lavender}95.43 & \cellcolor{Lavender}86.02 \\
\hline
\multirow{3}{*}{AdaptConv \cite{adapconv}}          & \cellcolor{Apricot}512  & \cellcolor{Apricot}90.45 & \cellcolor{Apricot}70.25 & \cellcolor{Apricot}96.01 & \cellcolor{Apricot}80.60 \\
                                                    & 1024 & 90.69 & 75.26 & 94.92 & 84.40 \\
                                                    & \cellcolor{Lavender}2048 & \cellcolor{Lavender}90.97 & \cellcolor{Lavender}75.08 & \cellcolor{Lavender}95.05 & \cellcolor{Lavender}84.72 \\ 
\hline
\multirow{3}{*}{PAConv \cite{paconv}}               & \cellcolor{Apricot}512  & \cellcolor{Apricot}91.97 & \cellcolor{Apricot}78.66 & \cellcolor{Apricot}95.66 & \cellcolor{Apricot}87.57 \\
                                                    & 1024 & 90.34 & 74.31 & 94.54 & 83.16 \\
                                                    & \cellcolor{Lavender}2048 & \cellcolor{Lavender}92.20 & \cellcolor{Lavender}70.59 & \cellcolor{Lavender}95.81 & \cellcolor{Lavender}79.18 \\ 
\hline
\multirow{3}{*}{3DMedPT}                        & \cellcolor{Apricot}512 & \cellcolor{Apricot}\textbf{94.82} & \cellcolor{Apricot}81.80 & \cellcolor{Apricot}\textbf{97.29} & \cellcolor{Apricot}89.25  \\
                                                & 1024 & 94.76 & \textbf{82.39} & 97.25 & \textbf{89.71} \\
                                                & \cellcolor{Lavender}2048 & \cellcolor{Lavender}93.52 & \cellcolor{Lavender}80.13 & \cellcolor{Lavender}96.59 & \cellcolor{Lavender}88.69 \\
\bottomrule
\end{tabular}
\end{table}

\noindent \textbf{ShapeNetPart.}
ShapeNetPart \cite{shapenet} contains 16,880 3D samples with 14,006 training and 2874 testing data, grouped into 16 classes with totally 50 part annotation labels.
We sample \cellcolor{Lavender}2048 points from each object and use mean intersection over union averaged across 16 classes (cls. mIoU) as the evaluation metric.
Table \ref{tab:shapenet} presents detailed per-class results and the overall cls. mIoU of our DGCNN backbone.
We achieve the overall value of 84.3\%, which is 0.4\% lower than PAConv but 1.1\% higher than AdaptConv.
Although we cannot achieve the best result, our model shows great performance in the class-wise segmentation results where we achieve the best in cap and mug.
Considering the performance gaps among ours, PAConv and AdaptConv on the IntrA segmentation task, we claim that our model can generalize well to general datasets.
Qualitative results are reported in the supplementary.

\begin{table*}
\small
\caption{Segmentation results of different methods on ShapeNetPart.} \label{tab:shapenet}
\begin{tabular}{l|@{ }c@{ }|@{ }c@{ }@{ }c@{ }@{ }c@{ }@{ }c@{ }@{ }c@{ }@{ }c@{ }@{ }c@{ }@{ }c@{ }@{ }c@{ }@{ }c@{ }@{ }c@{ }@{ }c@{ }@{ }c@{ }@{ }c@{ }@{ }c@{ }@{ }c@{ }}
\hline
\multirow{2}{*}{Method} & cls. & air   & \multirow{2}{*}{bag} & \multirow{2}{*}{cap} & \multirow{2}{*}{car} & \multirow{2}{*}{chair} & ear   & \multirow{2}{*}{guitar} & \multirow{2}{*}{knife} & \multirow{2}{*}{lamp} & \multirow{2}{*}{laptop} & motor & \multirow{2}{*}{mug} & \multirow{2}{*}{pistol} & \multirow{2}{*}{rocket} & skate & \multirow{2}{*}{table}  \\
&mIoU & plane & & & & & phone & & & & & bike & & & & board & \\ 
\hline
PointNet \cite{qi2017pointnet}    & 80.4 & 83.4 & 78.7 & 82.5 & 74.9 & 89.6 & 73.0 & 91.5 & 85.9 & 80.8 
                                       & 95.3 & 65.2 & 93.0 & 81.2 & 57.9 & 72.8 & 80.6 \\
PointNet++ \cite{qi2017pointnet2} & 81.9 & 82.4 & 79.0 & 87.7 & 77.3 & 90.8 & 71.8 & 91.0 & 85.9 & 83.7 
                                       & 95.3 & 71.6  & 94.1 & 81.3 & 58.7 & 76.4 & 82.6 \\
PointCNN \cite{pointcnn}   & 84.6 & 84.1 & \textbf{86.5} & 86.0 & 80.8 & 90.6 & 79.7 & \textbf{92.3} &                                      88.4 & 85.3 & 96.1 & 77.2 & 95.2 & 84.2 & 64.2 & 80.0 & 83.0 \\
DGCNN \cite{wang2019dynamic}    & 82.3 & 84.0 & 83.4 & 86.7 & 77.8 & 90.6 & 74.7 & 91.2 & 87.5 & 82.8 & 95.7
                                & 70.8 & 94.6 & 81.1 & 63.5 & 74.5 & 82.6 \\
KPConv \cite{thomas2019kpconv}  & \textbf{85.0} & 83.8 & 86.1 & 88.2 & \textbf{81.6} & 91.0 & 80.1 & 92.1 & 87.8 
                                & 82.2 & 96.2 & \textbf{77.9} & 95.7 & \textbf{86.8} & \textbf{65.3} &\textbf{81.7} & 83.6 \\
PointASNL \cite{pointasnl}    & 83.4 & 84.1 & 84.7 & 87.9 & 79.7 & 92.2 & 73.7 & 91.0 & 87.2 & 84.2                               & 95.8 & 74.4 & 95.2 & 81.0 & 63.0 & 76.3 & 83.2 \\
RS-CNN \cite{rsconv}          & 84.0 & 83.5 & 84.8 & 88.8 & 79.6 & 91.2 & 81.1 & 91.6 
                              & 88.4 & 86.0 & 96.0 & 73.7 & 94.1 & 83.4 & 60.5 & 77.7 & 83.6 \\
PCT \cite{guo2020pct}         & 83.1 & \textbf{85.0} & 82.4 & 89.0 & 81.2 & \textbf{91.9} & 71.5 
                              & 91.3 & 88.1 & \textbf{86.3} & 95.8 & 64.6 & 95.8 & 83.6 & 62.2 & 77.6 & 83.7 \\
PAConv \cite{paconv}          & 84.6 & 84.3 & 85.0 & 90.4 & 79.7 & 90.6 & 80.8 & 92.0 & \textbf{88.7} & 82.2 
                              & 95.9 & 73.9 & 94.7 & 84.7 & 65.9 & 81.4 & 84.0 \\
AdaptConv \cite{adapconv}     & 83.4 & 84.8 & 81.2 & 85.7 & 79.7 & 91.2 & 80.9 & 91.9 & 88.6 & 84.8 
                              & \textbf{96.2} & 70.7 & 94.9 & 82.3 & 61.0 & 75.9 & \textbf{84.2} \\ 
\hline
3DMedPT                       & 84.3 & 81.2 & 86.0 & \textbf{91.7} & 79.6 & 90.1 & \textbf{81.2} & 91.9 & 88.5 
                              & 84.8 & 96.0 & 72.3 & \textbf{95.8} & 83.2 & 64.6 & 78.2 & 83.8 \\
\hline
\end{tabular}
\end{table*}

\subsection{Ablation Study} \label{ablation}
In this section, we first examined the contribution of positional embeddings, and then investigated our model robustness on noise and compared the efficiency with some typical methods from Table \ref{intra_cls}.
Unless specified, all experiments are conducted on IntrA dataset for the 3D object classification with the first fold as the testing set and the others as the training set, and 1024 points are sampled for fast computation.

\noindent \textbf{Positional Embeddings.}
To investigate the effectiveness of the relative positional embedding, different models are established and F1 scores are averaged across all folds and reported in Table \ref{table:ablation}.
Model A is the design where no positional bias is introduced in the attention block, and we examine the cases when positional embedding is shared at all three positions individually (models B $\rightarrow$ D).
More investigations are done when positional embedding is introduced at any two positions (model E $\rightarrow$ G) or query\&key\&value (model H).


We can see from Table \ref{table:ablation} that without embedding position information, our model can still achieve a reasonable performance due to the design of our local context augmentation module.
Besides, introducing positional bias does improve the model performance when comparing model A with any others.
We find that introducing positional bias terms to all positions (model H) gives us the best result with the F1 score of 0.936 across all 5 folds, which indicates that more accurate positional information is learned via learning mapping in all query, key and value positions.

\begin{table}
\small
\centering
\caption{Positional embeddings in 3DMedPT evaluated on F1-score.
$\text{RPE}_{k}$, $\text{RPE}_{q}$ and $\text{RPE}_{v}$ indicate relative positional embeddings at the key, query and value.
}\label{table:ablation}
\begin{tabular}{l|ccc|c}
\toprule
& $\text{RPE}_{q}$ & $\text{RPE}_{k}$ & $\text{RPE}_{v}$ & F1  \\ 
\hline
A &              &              &              & 0.905 \\
B & $\checkmark$ &              &              & 0.924 \\
C &              & $\checkmark$ &              & 0.915 \\
D &              &              & $\checkmark$ & 0.921 \\
E & $\checkmark$ &              & $\checkmark$ & 0.920 \\
F &              & $\checkmark$ & $\checkmark$ & 0.917 \\
G & $\checkmark$ & $\checkmark$ &              & 0.928 \\
H & $\checkmark$ & $\checkmark$ & $\checkmark$ & 0.936 \\
\bottomrule
\end{tabular}
\end{table}

\noindent \textbf{Model Efficiency.}
Computational costs of different models compared with 3DMedPT were explored in terms of the model size and processing speed.
As shown in Table \ref{tab:model_eff}, we achieved the highest performance with the smallest model size which only contains 1.54M model parameters, processing 843 examples per second.
Although the processing speed of PointNet \cite{qi2017pointnet} is the fastest, it cannot perform as well as other models at a lower computational speed.
Moreover, our Transformer counterpart PCT \cite{guo2020pct} requires more trainable parameters to achieve a relatively good performance, with a slower processing speed than ours.


\begin{table}
\centering
\small
\caption{
Model complexity of 3DMedPT for the IntrA classification, where model parameters are in the unit of millions and throughput is reported in examples per second.}
\label{tab:model_eff}
\begin{tabular}{l|cc|c}
\hline
Method            & \#Params  & Throughput & F1 \\
\hline
PointNet++ \cite{qi2017pointnet2}       & 1.75M            & 245 ex/s               & 0.769    \\
DGCNN \cite{wang2019dynamic}            & 1.81M            & 365 ex/s               & 0.738    \\
PCT \cite{guo2020pct}                   & 2.73M            & 471 ex/s               & 0.872    \\
AdaptConv \cite{adapconv}               & 1.76M            & 230 ex/s               & 0.806    \\
PAConv \cite{paconv}                    & 2.32M            & 507 ex/s               & 0.866    \\
3DMedPT                        & \textbf{1.54}M   & \textbf{843} ex/s      & \textbf{0.922}    \\
\hline
\end{tabular}
\end{table}

\begin{figure}
\centering
\includegraphics[width=\linewidth]{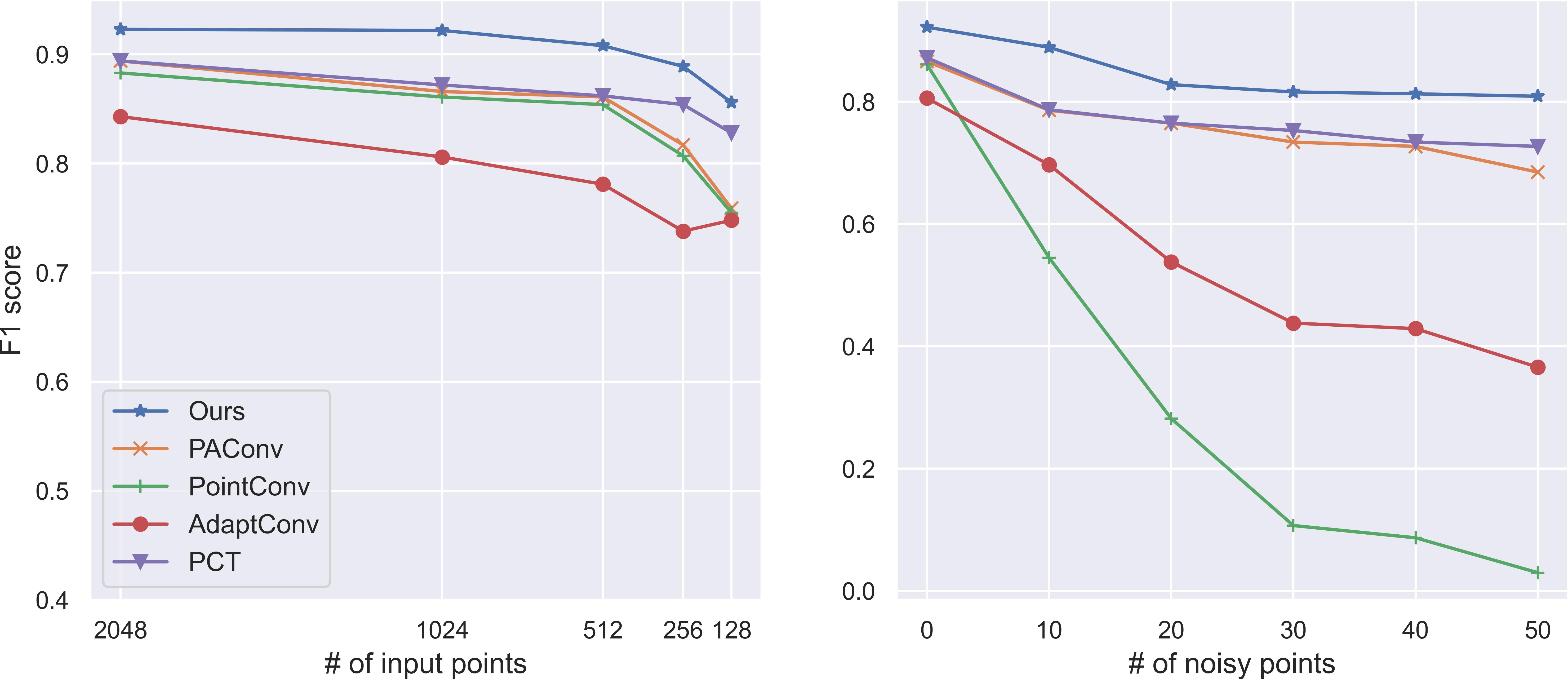}
\caption{
\textbf{Left}: Comparison on different numbers of input points.
\textbf{Right}: Comparison on different numbers of noisy points.
} \label{fig:noise}
\end{figure}

\noindent \textbf{Robustness Analysis.}
We demonstrated our model robustness to the point density by using sparser points as the network input from \cellcolor{Lavender}2048 to 128 points.
We compared our results with several works in Fig. \ref{fig:noise} (left), where all networks were trained with 1k points on IntrA.
The absolute dropping difference from \cellcolor{Lavender}2048 to 128 points of our method is 6.7\%, which is the same as our Transformer counterpart PCT, while we reached the best F1 score on all experiments with different numbers of input points.
For the noise resistance investigation, we introduced different numbers of noisy 3D points with random positions during model testing following \cite{pointasnl}.
As can be seen from Fig. \ref{fig:noise} (right), 3DMedPT is more robust to noise compared with some latest works PAConv \cite{paconv} and AdaptConv \cite{adapconv} under all testing environments.
The absolute difference between no noise and 50 noisy points for our model is 11.3\%, which is smaller than PCT with the value of 14.5\%, presenting our excellent robustness ability to the noise.

\section{Conclusion}
In this work, we propose a Transformer network for 3D medical point cloud analysis, namely 3DMedPT, which can model long-range dependencies of global contents via the convolutional operation introduced at query to summarize local feature responses, and local context interactions based on lambda attention modified with local context augmentation.
Variant relative positional information for query, key and value is encoded to capture the complex structure of medical data.
Global interactions between features are obtained from channel space where multiple graphs are constructed to model diverse graph states, improving the expressiveness of feature information.
Our model performs the best in 3D medical object classification and part segmentation tasks.
Moreover, extensive analyses on general 3D point cloud datasets have validated the good generalization ability of our model. In future, we will extend our approach for large-scale point cloud processing.

{\small
\bibliographystyle{ieee_fullname}
\bibliography{ms}
}

\end{document}


\title{Supplementary Materials}

\author{}
\maketitle
\ificcvfinal\thispagestyle{empty}\fi

\section{Training Details}
We used PyTorch on one 11GB RTX2080Ti GPU to implement our model for IntrA \cite{yang2020intra}, ModelNet40 \cite{xie2016deepshape}, and ShapeNetPart \cite{shapenet}.

\noindent \textbf{IntrA Classification.}
Data augmentation was applied during the training stage, where random point translation in [-0.2, 0.2] and rescaling in [0.67, 1.5] were adopted.
We set the training batch size to 32 and testing batch size to 16.
We trained the model for 250 epochs with Adam as the optimizer, where the learning rate was set to 1e-2 with momentum of 0.9 and weight decay of 1e-4.
Cosine annealing was applied to reschedule the learning rate for each epoch.
We used binary cross entropy as the loss function for the classification.

\noindent \textbf{IntrA Segmentation.}
Data augmentation utilized for segmentation task follows the same as classification, while both the training and testing batch size were set to 8.
We trained the model with 400 epochs using Adam, where the learning rate was set to 1e-3 with weight decay value of 1e-4, and we applied cosine annealing to adjust the learning rate.
The same loss function as the classification task was used.

\noindent \textbf{ModelNet40.}
We used the standard data split as \cite{qi2017pointnet} where 9,843 samples were used for training and the remaining 2,468 were taken as the testing samples. 
We also applied FPS \cite{qi2017pointnet2} to downsample the number of points during network training.
The same data augmentation technique is applied .
The training batch size was set to 32 and the testing batch size was set to 16.
The model was trained for 250 epochs with initial learning rate set to 1e-3. 
Adam was used as the optimizer with cosine annealing, and weight decay and momentum were set to 1e-4 and 0.9, respectively.
The multi-class cross entropy was adopted as the loss function for the classifications of 40 objects.

\noindent \textbf{ShapeNetPart.}
The data augmentation techniques remain the same as the above experiments on ModelNet40.
The total training epochs was set to 400 with 32 training batches and 16 testing batches.
The learning rate was initialized to 1e-3, and Adam was selected as the model optimizer with a weight decay of 1e-4.
We also applied the cosine annealing scheme to dynamically adjust the learning rate, with multi-class cross entropy adopted as the loss function.
The loss function to guide the model training was the same as ModelNet40.

\section{Effectiveness of MGR}
To examine the effectiveness of MGR module, we used the first fold for testing and all the others for training with 1024 input points.
As can be seen in Table \ref{tab:mgr}, experimental results of model with no MGR, with only one graph and with multiple graphs are reported.
When MGR is ignored during the training stage, the final result is only 0.884, which is 4.3\% smaller than the desired design where multiple graphs are used.
Besides, with only one graph, our model can give a good performance with F1 score of 0.910.

\begin{table}
\centering
\caption{Model performance on IntrA classification with different configurations of MGR, where MGR-1 indicates only one graph is constructed.}
\label{tab:mgr}
\begin{tabular}{l|c}
\toprule
Method          &   F1   \\
\hline
w/o MGR         &  0.884 \\
MGR-1           &  0.910 \\
MGR             &  \textbf{0.922} \\
\bottomrule
\end{tabular}
\end{table}

\section{Depth of Architecture}
We investigated the impact of model depth on the performance when handling the classification task of the IntrA dataset.
We used the first fold to testing and all others for training with 1024 input points.
Specifically, we vary the number of attention layers and fix the MGR layer at the end.
The F1 scores and model parameters of four different configurations are reported in Table \ref{tab:layers}. 
It can be seen that more numbers of layers mean slower processing speed where the 1-layered form can process 1229 examples per second. The best model configuration is empirically found to be the 2-layered one, which is the one that we applied for the blood vessel and aneurysm classification. For the trade-off between the model efficiency and the classification accuracy, we prefer the higher F1 score for more accurate disease diagnosis.

\begin{table}
\centering
\caption{Model performance comparisons among four different layer settings, where \#Layers denotes the number of attention modules before the MGR module.} \label{tab:layers}
\begin{tabular}{l|cc} 
\toprule
\#Layers & F1 & Throughout  \\ 
\hline
1     & 0.902               & \textbf{1229} ex/s        \\
2     & \textbf{0.922}      & 843 ex/s                  \\
3     & 0.912               & 508 ex/s                  \\
4     & 0.901               & 326 ex/s                  \\
\bottomrule
\end{tabular}
\end{table}

\section{Hyperparameter Search}
To optimize the attention module design, we examined the influence of two hyperparameters, i.e., the number of multi-query heads $h$ and query feature $C_k$.
F1 scores of models with different hyperparameter combinations with 1024 points are reported in Table \ref{tab:hk}, where we used the first fold as testing set and the others as training sets.
It can be seen that the best F1 score of 0.922 is achieved when $C_k=32$ and $h=8$.

\begin{table}
\centering
\caption{Model performances of different hyperparameter choices for the optimal attention block design.}\label{tab:hk}
\begin{tabular}{>{\centering}p{.05\linewidth}|l|cccc} 
\toprule
\multicolumn{2}{l|}{\multirow{2}{*}{F1 Score}} & \multicolumn{4}{c}{$C_k$}      \\ 
\cline{3-6}
\multicolumn{2}{l|}{}                                  & 8     & 16     & 32    & 48    \\ 
\hline
\multirow{4}{*}{h} & 8                                 & 0.913 & 0.912 & \textbf{0.922} & 0.916  \\
                   & 16                                & 0.912 & 0.913 & 0.921 & 0.919   \\
                   & 32                                & 0.919 & 0.918 & 0.921 & 0.916  \\
                   & 64                                & 0.915 & 0.910 & 0.916 & 0.904  \\
\bottomrule
\end{tabular}
\end{table}

\section{Normalization Functions}
We further investigated the behavior difference when using different normalization functions.
Common choices for normalization such as sigmoid and Tanh were applied to check their effectiveness.
Extensive experiments are implemented on IntrA classification task, and the classification results are reported in Table \ref{tab:norm}.
It suggests that summarizing the contextual information by softmax is the best choice.
\begin{table}
\centering
\caption{Model performance on IntrA classification with different normalization functions.}
\label{tab:norm}
\begin{tabular}{l|c}
\toprule
Norm. function & F1  \\
\hline
w/o norm       &  0.887 \\
sigmoid        &  0.915 \\
max(0, Tanh)   &  0.913 \\
softmax        &  \textbf{0.922} \\
\bottomrule
\end{tabular}
\end{table}

\section{More Results on Medical Datasets}
While we focused on medical point clouds, we followed PointNet \cite{qi2017pointnet} to implement a sanity check on our method with experiments on other medical data: RetinalOCT \cite{retinal} and AdrenalMNIST3D \cite{medmnistv2}, where shape analysis is critical for disease classification in these two datasets.

\noindent \textbf{RetinalOCT.}
RetinalOCT involves 2D gray-scale images of retinal diseases, which is comprised of 4 diagnosis categories with 108,318 training and 1,000 testing samples from 633 patients, leading to a multi-class classification task.
To make it compatible for training with our model, we convert 2D image pixels to 2D points.
Specifically, we firstly resize each image to 256x256 and use Sobel filters to detect the edges of each image, with $(x, y)$ assigned by (row, col) of each pixel and $z = 0$.
The number of points for each data sample is fixed and determined by the average value of all point sets obtained from all images.
During training, we normalize each point set to a unit cube of [-1, 1] and use data augmentation following IntrA.

In Table \ref{table:oct}, we compared with ResNet-18/50 baselines \cite{skip} with 3D convolutions and an open-source AutoML tool (AutoKeras \cite{jin2019auto}).
The highest testing accuracy is achieved by \cite{retinal} where InceptionV3 \cite{inceptionv3} is applied and pre-trained on ImageNet.
Although there are noticeable performance gaps between our approach and CNN-based methods, we argue that it is the information loss due to the data type conversion (i.e., image resizing and ignorance of pixel intensity) that causes the performance difference.

\begin{table}
\centering
\caption{Classification results on RetinalOCT with different methods.
} \label{table:oct}
\begin{tabular}{l|ccc} 
\toprule
Method         & Input Type      & Acc. (\%)  \\ 
\hline
Pre-trained InceptionV3 \cite{retinal}  & pixels & 96.6 \\
ResNet-18 \cite{skip}           & pixels & 95.8 \\
ResNet-50 \cite{skip}           & pixels & 96.1 \\
AutoKeras \cite{jin2019auto}    & pixels & 96.3 \\
\hline
PointNet \cite{qi2017pointnet}  & points & 86.5 \\
DGCNN \cite{wang2019dynamic}    & points & 87.9 \\
PCT \cite{guo2020pct}           & points & 87.2 \\
3DMedPT                         & points & 90.9 \\
\bottomrule
\end{tabular}
\end{table}

\noindent \textbf{AdrenalMNIST3D.}
AdrenalMNIST3D is a new 3D dataset collected from abdominal computed tomography for shape classification, consisting of shape masks with the size of 28x28x28 from 1,584 left and right adrenal glands from 792 patients.
We first pre-process the volumetric images to transfer them to point clouds.
To eliminate the advantages brought by pre-processing steps, we simply follow the steps from PointNet \cite{qi2017pointnet}, where we only added voxels to our point set with intensity is greater than 128.
The size of point set is 256 and normalized to a unit cube within the range of [-1, 1].
Point positions are represented by ($x,y,z$) coordinates of voxels.

As reported in Table \ref{table:AdrenalMNIST3D}, we compared several point-based methods with well-engineered CNNs.
It can be seen that the best model is sill the well-engineered CNN-based model, while our model can achieve a reasonable performance and also the accuracy of our work is quite near AutoKeras when converting 3D images to points in a simple fashion.

\begin{table}
\centering
\caption{Classification results on AdrenalMNIST3D with different methods.
} \label{table:AdrenalMNIST3D}
\begin{tabular}{l|ccc} 
\toprule
Method         & Input Type      & Acc. (\%)  \\ 
\hline
ResNet-18 \cite{skip}           & voxels & 82.7 \\
ResNet-50 \cite{skip}           & voxels & 82.8 \\
AutoKeras \cite{jin2019auto}    & voxels & 80.4 \\
\hline
PointNet \cite{qi2017pointnet}  & points & 76.5 \\
DGCNN \cite{wang2019dynamic}    & points & 77.1 \\
PCT \cite{guo2020pct}           & points & 76.8 \\
3DMedPT                         & points & 79.1 \\
\bottomrule
\end{tabular}
\end{table}

\section{Qualitative Results}
\begin{figure*}
\centering\includegraphics[width=0.88\textwidth]{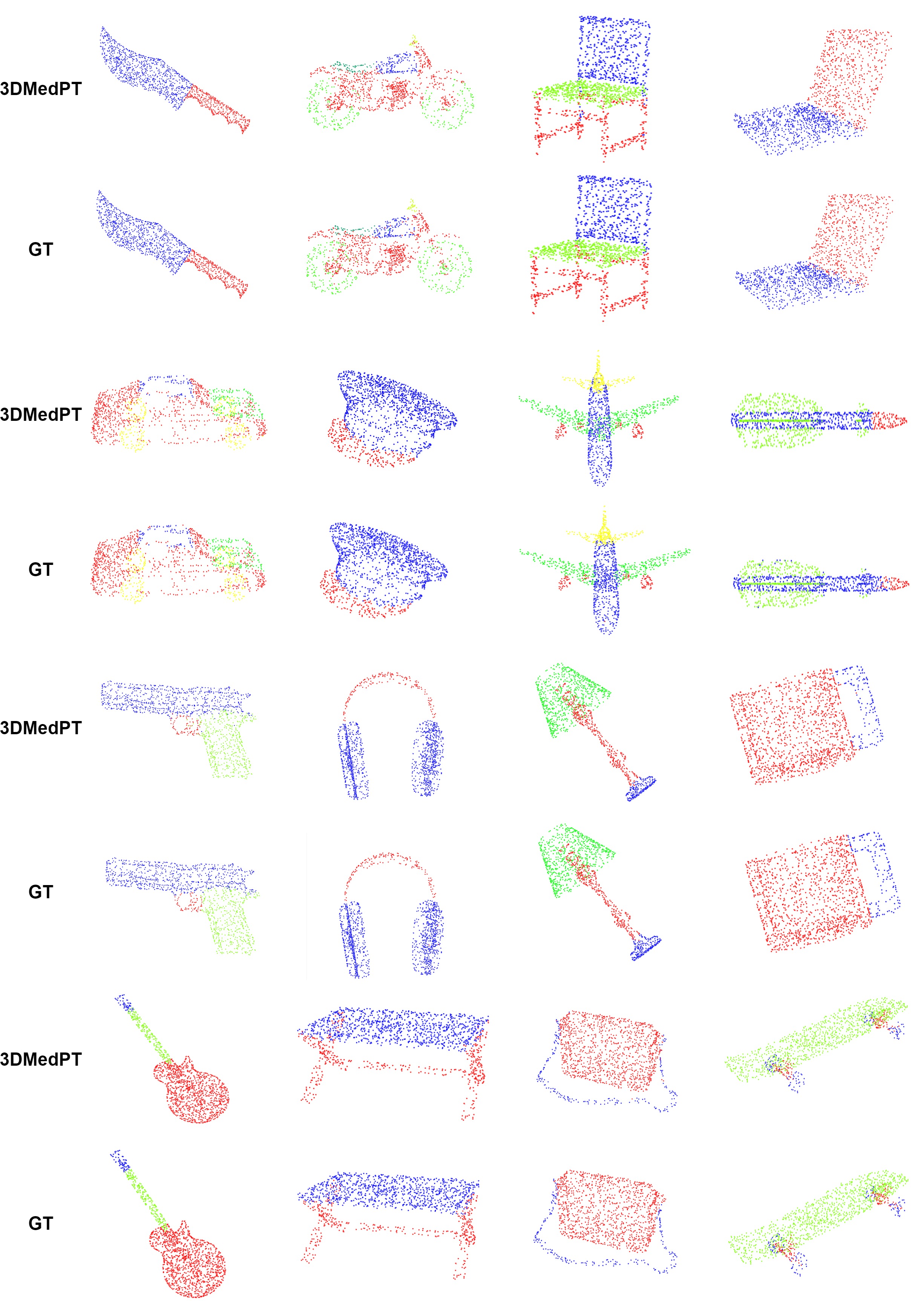}
\caption{
Segmentation comparisons between groundtruth (GT) annotations and the outputs generated from the 3DMedPT on ShapeNetPart dataset.
} \label{fig:shapenet}
\end{figure*}

\noindent \textbf{ShapeNetPart.}
Object part segmentation results of our model are shown in Fig. \ref{fig:shapenet}, along with the groundtruth examples displayed for references. We tested our model performance on each class and reported the visual outputs across all 16 classes. It can be seen that the segmented samples on mug, laptop, chair and laptop show almost no difference with the original labels, while there are still defects on segmentation results for motorbikes and rockets.

\begin{figure*}
\centering\includegraphics[width=1\textwidth]{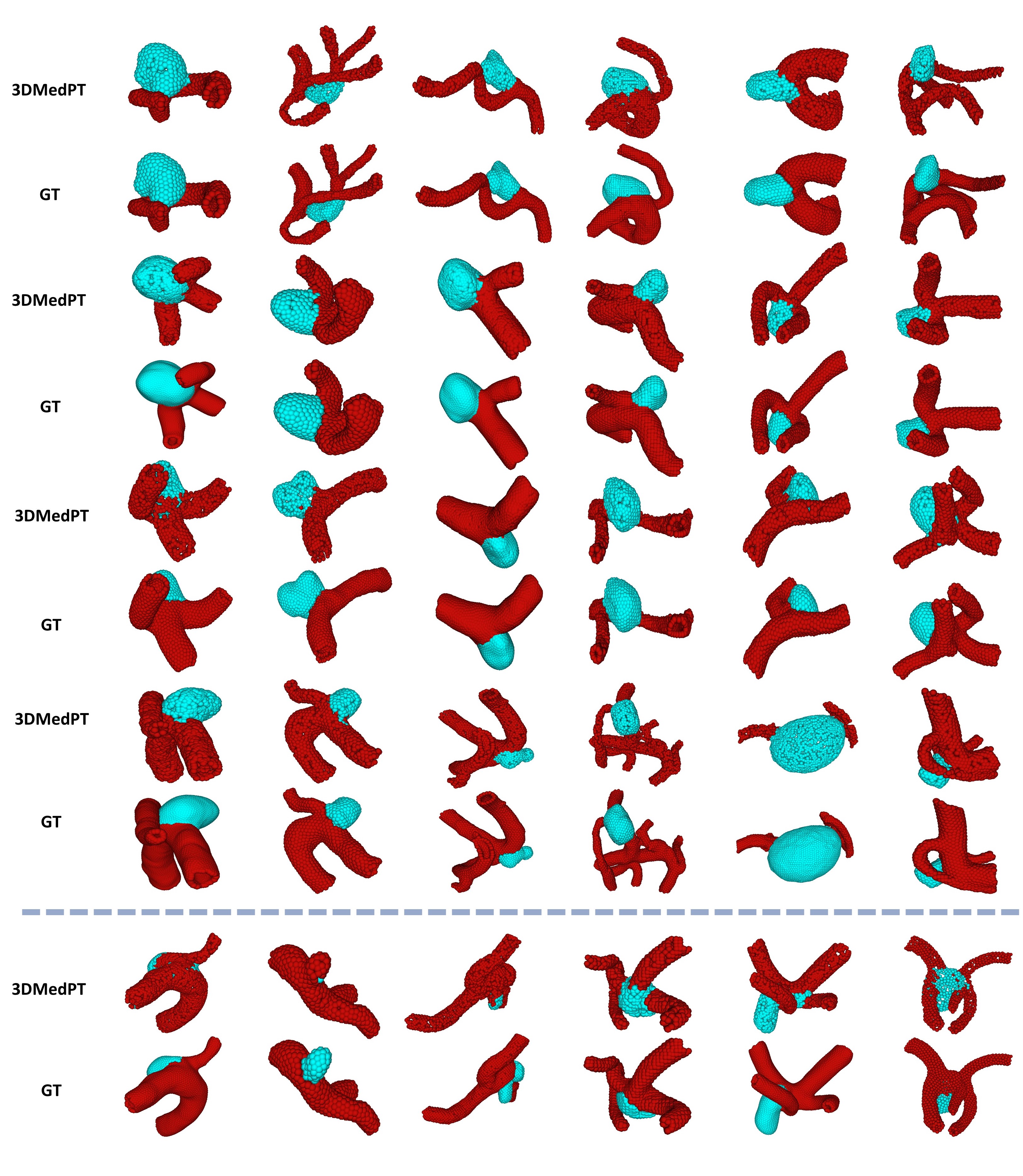}
\caption{
Segmentation comparisons between groundtruth (GT) annotations and the outputs generated from the 3DMedPT on IntrA dataset. Good segmentation results are shown above the red line and some failed cases are shown below the red line.
} \label{fig:pics}
\end{figure*}

\noindent \textbf{IntrA.}
More visual results on the segmentation task of IntrA dataset are shown in Fig. \ref{fig:pics} based on our best model, which are compared to the corresponding groundtruth annotations for comprehensiveness.
As shown in Fig. \ref{fig:pics}, our method achieves fairly precise segmentation results on most cases, however, few undesired results might appear especially when the size ratio of aneurysms becomes smaller than the healthy blood vessels, or when the 3D structure topology becomes complicated.

{\small
\bibliographystyle{ieee_fullname}
\bibliography{ms}
}